\documentclass[prd,amsmath,amssymb,longbibliography,superscriptaddress,twocolumn,nofootinbib,10pt]{revtex4}

\usepackage{graphicx}
\usepackage{dcolumn}
\usepackage{bm}
\usepackage{amssymb}
\usepackage{latexsym}
\usepackage{booktabs}
\usepackage{amsmath}
\usepackage{multirow}
\usepackage[colorlinks=true, linkcolor=blue, citecolor=blue]{hyperref}

\def\aj{Astron. J.}
\def\aap{Astron. Astrophys.}
\def\mnras{Mon. Not. Roy. Astron. Soc.}
\def\apjl{Astrophys. J. Lett.}
\def\jcap{JCAP}

\def\apjs{Astrophys. J., Suppl. Ser.}

\newcommand{\Mpc}{\mathrm{~km~s^{-1}~Mpc^{-1}}}
\begin{document}

\title{Model independent calibration for sound horizon combining observations of supernovae and transversal BAO measurements}

\author{Tonghua Liu}
\affiliation{School of Physics and Optoelectronic, Yangtze University, Jingzhou 434023, China;}
\author{Xinyi Zhong}
\affiliation{School of Physics and Optoelectronic, Yangtze University, Jingzhou 434023, China;}
\author{Jieci Wang}
\email{jcwang@hunnu.edu.cn}
\affiliation{Department of Physics, and Collaborative Innovation Center for Quantum Effects and Applications, Hunan Normal University, Changsha 410081, China;}
\author{Marek Biesiada}
\email{Marek.Biesiada@ncbj.gov.pl}
\affiliation{National Centre for Nuclear Research, Pasteura 7, PL-02-093 Warsaw, Poland;}

\begin{abstract}
The sound horizon is a key theoretical prediction of the cosmological model that depends on the speed of sound and the rate of expansion in the early universe, before matter and radiation decoupled.
The standard ruler for low redshift calibration of baryon acoustic oscillations (BAOs) is a direct measurement that would exist even if the standard cosmological model and the standard assumptions of early physics did not. We propose a new model-independent method to calibrate sound horizon $r^h_s$ (relative standard ruler) by using the latest observations of SNe Ia and transversal BAO measurements. The final result reports  $r_s^{h}=107.10^{+1.36}_{-1.32}$ $Mpc/h$ in the framework of the Pantheon dataset. This result changes to $r_s^{h}=105.63^{+1.33}_{-1.31}$ $Mpc/h$ when uses Pantheon+ dataset.  Note that even without an estimate of dimensionless Hubble constant $h$, the combination of BAO and SNe Ia datasets already constrain the low-redshift standard ruler scale $r_s^{h}$ at the $\sim1.26\%$ level.  More importantly, it is interesting to find that most of the $r_s^{h}$ obtained at high redshifts have a larger value (9 out of 15 results are larger than the result obtained by combining all BAOs).  This finding may give us a better understanding of the discordance between the data sets or Hubble tension or reveal new physics beyond the standard model.

\end{abstract}

\maketitle

%\bigskip
\section{Introduction}
In the past few decades, observations of Type Ia supernovae (SNe Ia) have revealed that the expansion of the Universe is accelerating \cite{1998AJ....116.1009R,1999ApJ...517..565P}. Based on Einstein's theory of General Relativity (GR) \cite{1984ucp..book.....W}, and assuming that the Universe is homogeneous and isotropic on large scales \citep{1972gcpa.book.....W}, it is generally accepted that the so called  cosmological constant $\Lambda$ plus cold dark matter ($\Lambda$CDM) model correctly describes our Universe, where the cosmological constant term $\Lambda$ modeling dark energy drive the accelerated expansion of the Universe. However, with the improved precision of the latest surveys, the standard cosmological model is being severely tested, such as, the Hubble tension \citep{2017NatAs...1E.169F,2021APh...13102605D}. Assuming the $\Lambda$CDM model, Planck Collaboration yields a low value on Hubble constant, $H_0=67.36 \pm 0.54 \Mpc$, from cosmic microwave background (CMB) anisotropy (both temperature and polarization) data \citep{2020A&A...641A...6P}. This result suffer challenge by some compatibility tests at low redshifts. In particular, the \textit{Supernova $H_0$ for the Equation of State} collaboration (SH0ES) \citep{2019ApJ...876...85R} gives the low redshift measurement result on the Hubble constant, $H_0=74.03\pm 1.42 \Mpc$, through a local distance ladder approach from the Cepheid variable stars. The local measurement of $H_0$ is model independent because it does not depend on cosmological assumptions. This inconsistency on $H_0$ is called Hubble tension, which possible explanations for this tension are unknown systematic errors in one or both observations, or problems on the standard model with unrevealed new physics.
(more works on Hubble tension please see the references \cite{2019PhRvL.122v1301P,2019PhRvL.122f1105F,2021ApJ...912..150D,2022A&A...668A..51L} and  references therein). The search for other high-precision and accuracy astronomical observations may be an important way to relieve or even determine Hubble tension.

The baryon acoustic oscillation (BAO) leaves measurable signatures in the distribution of galaxies and is probably the most easily understood standard scale (ruler) in the Universe. For instance, the Sloan Digital Sky Survey (SDSS) has increased the three-dimensional catalog of galaxies \cite{2018ApJS..235...42A} , making it possible to accurately measure BAO scales at various redshifts. Thus, probing the BAO scale at different times is
a powerful tool in constraining cosmology. However, it should be noted that the BAO acting standard ruler need to know the comoving length of the ruler, the sound horizon at radiation drag $r_s$ \cite{1998ApJ...496..605E,1996ApJ...471...30H}.
Without knowing the length of the ruler, BAO probes can only give relative measurements of the expansion history. The $r_s$ is usually calibrated at $z\approx 1100$ relying on the CMB observations. If one use BAO data with calibrated $r_s$ by CMB observations to measure $H_0$, in some sense, the constraint on $H_0$ is not completely independent on the CMB data \cite{2020SCPMA..6390402Z}.  An alternative approach is to combine BAO measurements with other low-redshift observations. The BAO signature can independent estimates of the the Hubble parameter $H(z)$ and the angular diameter distance $D^A(z)$ through the line-of-sight and transverse BAO modes, when one knows the length of the ruler, respectively \cite{2017MNRAS.470.2617A,2021PhRvD.103h3533A}. This means that if we calibrate the length of the ruler of  BAO from low redshift observations, we need to add both $H(z)$ and $D^A(z)$ data. Although astronomical observations such as cosmic chronometers can directly provide cosmological model-independent measurements of $H(z)$, and SNe Ia can restore the distance redshift relationship and provide the expansion history of the Universe, using these data together makes it impossible to determine the contribution of the various data to calibrating the length of the ruler in BAO.

Even relative measurements of the expansion history provided by observations of SNe Ia, combined with a measurement of the BAO signature, can impose constraints on the relative standard ruler $r_s^h$, which is the ruler length in units of $h^{-1}$Mpc. An absolute ruler length can be measured by multiplied dimensionless Hubble constant $h=H_0/100 \Mpc$. The importance of this scale is that it is a key theoretical prediction for cosmological models that depend on the speed of sound and the rate of expansion in the early Universe, before matter and radiation decoupled. However, this standard scale still exists even without the hypothetical cosmological model and early cosmology, which means that we can calibrate this standard ruler with astronomical observations at low redshifts. We refer to Refs \cite{2014PhRvL.113x1302H,2022PhRvD.105d3528G,2016JCAP...10..019B,2017JCAP...01..015L,2019ApJ...874....4A,2019MNRAS.486.2184M,2019MNRAS.486.5046W} for more papers about the sound horizon. Inspired by the above, this work will use the observational datasets of SNe Ia, combined with the measurement of transverse BAO, to calibrate the relative standard ruler $r_s^{h}$. This has several advantages. First, the whole work has nothing to do with cosmological models and the early universe; Second, we only calibrate the relative standard ruler, which has nothing to do with the Hubble constant; Third, the work only involves the SNe Ia and BAO datasets, and has nothing to do with other astronomical observations, which highlights the contribution of each datasets in the overall work.

This paper is organized as follows: in Section 2, we present the methodology of calibration of relative standard ruler, and the data used in this work. In Section 3, we show our results and discussion. Finally, the main conclusions are summarized in Section 4.

\section{ Methodology and  Data}\label{sec:data}
\subsection{Transverse BAO measurements}

BAOs are regular periodic density fluctuations of the visible baryonic matter in the Universe. The matter clustering properties of BAOs can also be used as a standard ruler for measuring cosmological distances, thus to constrain the cosmological parameters (especially the density of baryonic matter), and further understand the nature of the dark energy that causes the accelerated expansion of the Universe. However, BAO data can not obtain a constraint on $H_0$, it must be combined with other measurements in order to break the degeneracy between $H_0$ and $r_s$. The sound horizon at the drag epoch is given by
\begin{equation}
r_s=\int^{\infty}_{z_d}dzc_s(z)/H(z),\label{eq1}
\end{equation}
where $z_d$ being the redshift of the drag epoch, $c_s(z)$ is the sound speed of the photon-baryon fluid, and $H(z)$ is the Hubble parameter. The BAO angular scale $\theta_{BAO}(z)$ is related to the angular diameter distance acting as standard ruler and given by following
\begin{equation}
	\theta_{BAO}=\frac{r_s}{(1+z)D^A(z)},\label{eq2}
\end{equation}
where $D^A(z)$ is the angular diameter distance at the redshift $z$. The calibration of $r_s$ enables the utilization of the BAO angular scale $\theta_{BAO}$ to establish a precise correlation between angular diameter distance and redshift.
In this work, we consider the 15 transverse BAO angular scale measurements. These values were obtained using public data releases (DR) of the Sloan Digital Sky Survey (SDSS), namely: DR7, DR10, DR11, DR12, DR12Q (quasars), without assuming a fiducial cosmological model, which are summarized in Table 1 of \citet{2020MNRAS.497.2133N}.  It is important to note that because these lateral BAO measurements are performed using cosmology-independent methods, their errors are larger than those obtained using the fiducial cosmological method. Nevertheless, this dataset is larger than other BAO datasets such as (3D-BAO only provides seven line-of-sight and transverse measurements data points \cite{2021PhRvD.103h3533A}, the Dark Energy Spectroscopic Instrument only provides five data points \cite{2024arXiv240403002D}).

As we mentioned above, if the calibration of $r_s$ independent of cosmological models is to be achieved, then other low redshift observations need to be combined. Here we consider observations of SNe Ia. It is important to note that the distance provided by SNe Ia is relative distance. Therefore, we can only calibrate the length of the relative standard ruler $r_s^{h}=r_s*h$. If the value of the Hubble constant was the one measured locally,  it would yield a smaller value on $r_s$.
\begin{figure*}
\begin{center}
{\includegraphics[width=0.45\linewidth]{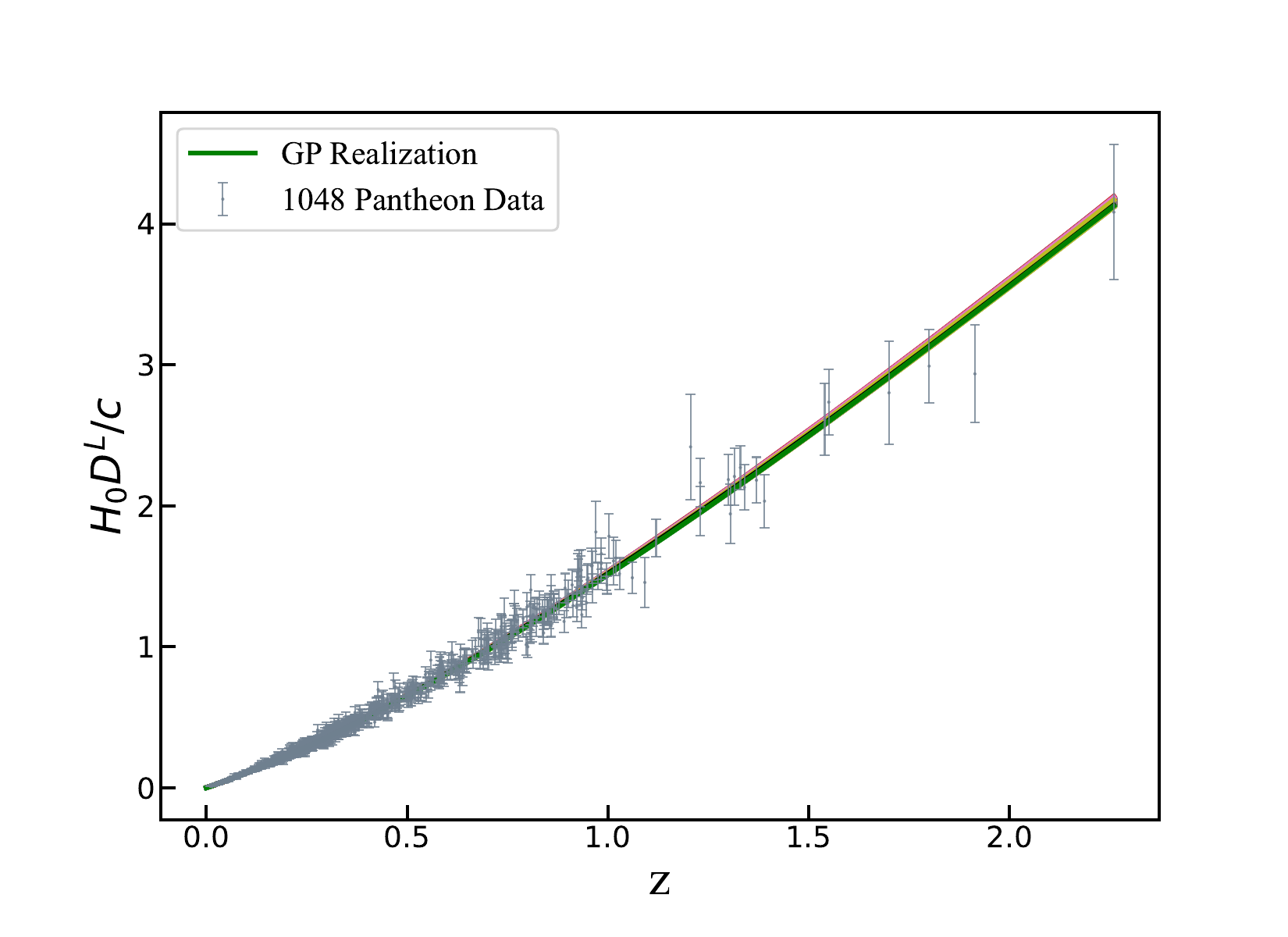}
\includegraphics[width=0.45\linewidth]{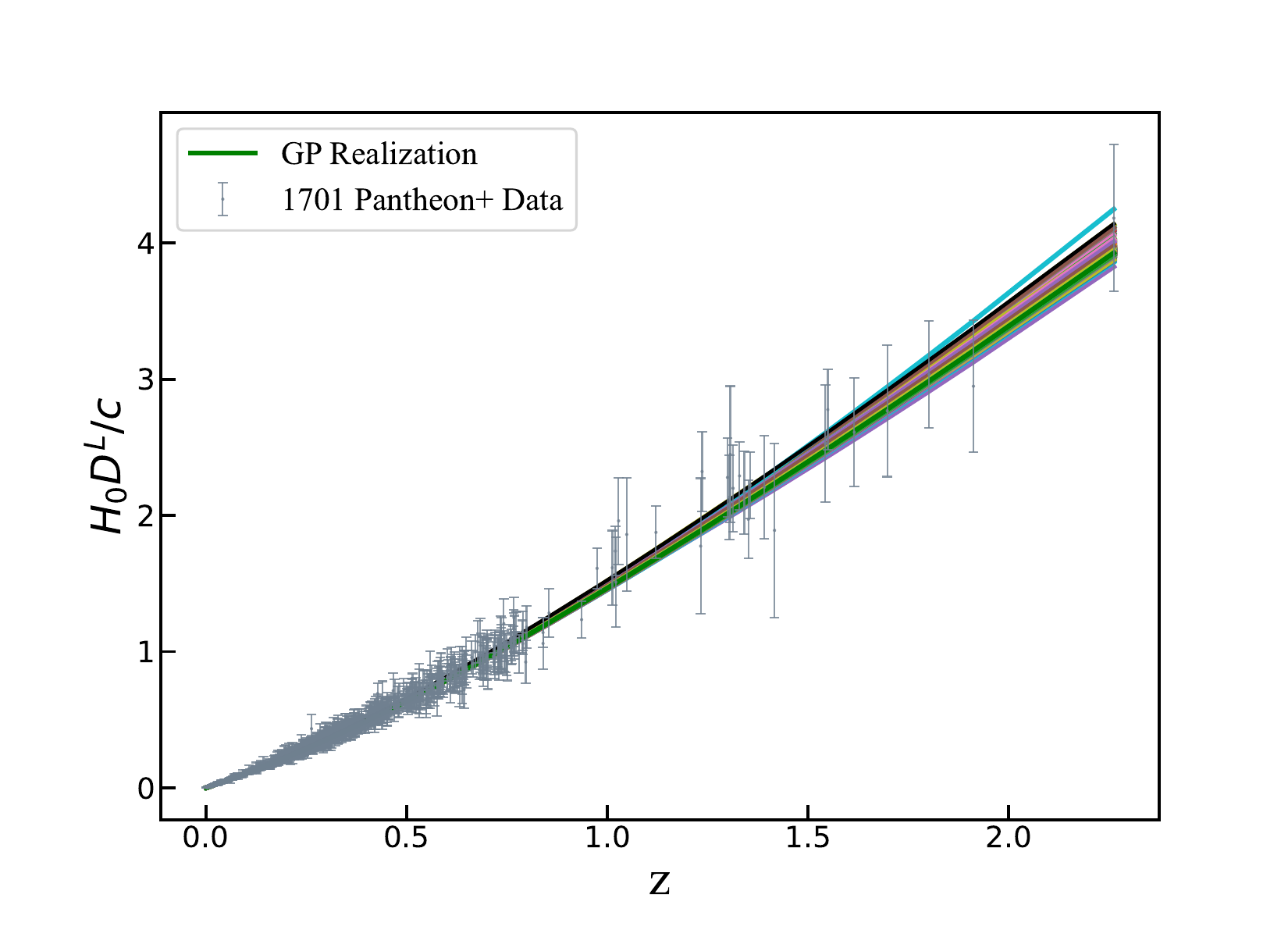}}
\end{center}
\caption{{\it Left panel:} The 1000 reconstructed unanchored distance $H_0D^L/c$ by using GPR with Pantheon dataset; {\it Right panel:} Similar to left figure, but for  Pantheon+ dataset. The dots with error bars represent observations of SNe Ia. }\label{fig:sn}
\end{figure*}

\subsection{Reconstructions of distance redshift relations from observations of SNe Ia}
As the most common transient source in the universe, the SNe Ia, by their very nature, can act as standard candles in the Universe and are regarded as powerful cosmological probes.
It was observations of SNe Ia that revealed that the universe is undergoing accelerated expansion \cite{1998AJ....116.1009R,1999ApJ...517..565P}. Previously, \citet{2018ApJ...859..101S} combined the subset of 279 Pan-STARRS1(PS1) ($0.03 < z < 0.68$) supernovae \citep{2014ApJ...795...44R,2014ApJ...795...45S} with the useful data of SNe Ia from SDSS, SNLS, and various low redshift and HST samples to form the largest combined sample of SNe Ia consisting of a total of 1048 SNe Ia ranging from $0.01<z<2.3$, which is known as the  ``Pantheon" sample. We refer to work \citep{2018ApJ...859..101S} for more details about the SNe Ia standardization process including the improvements of the PS1 SNe photometry, astrometry and calibration. Most recently, an updated sample named ``Pantheon+" was reported in work of \cite{2022ApJ...938..113S}, which consists of 1701 light curves of 1550 distinct SNe Ia spanning the redshift range $0.001\leq z\leq2.26$. This larger SNe Ia sample has a significant increase compared to the original Pantheon sample, especially at lower-redshifts.

However, while the SNe Ia datasets are rich enough, BAO data points are hard to match at the same redshift.
To combine the SNe Ia and BAO transverse measurements datasets, we generate samples of the unanchored luminosity distances $H_0D^L$ from the posteriors of the Pantheon and Pantheon+ datasets, respectively. We use the Gaussian process regression (GPR)~\citep{Holsclaw1,Keeley0,ShafKimLind,Aghamousa2017} to realize posterior sampling by using the code \texttt{GPHist}\footnote{https://github.com/dkirkby/gphist.} without assuming cosmological models \citep{GPHist}. GPR is a non-parametric regression method that models relationships between data based on Gaussian process models. GPR is performed in an infinite dimensional function space without overfitting problems and is a powerful tool for function reconstruction \citep{Keeley0}. Meanwhile, We use SNe datasets to generate a large number of function samples $\gamma(z) = \ln([H^{\rm fid}(z)/H_0]/[H(z))/H_0])$, and obtain the expansion histories $H(z)/H_0$. We refer to work \citep{2024MNRAS.528.1354L,2024ApJ...960..103L} for more details about using GPR to reconstruct expansion history from observations of SNe Ia.
With the reconstructed expansion history $H(z)/H_0$, the unanchored SN luminosity distances can be calculated by
\begin{equation}
    H_0 D^L = (1+z) \int^z_0 dz'/[H(z')/H_0]\ .
\end{equation}
The final reconstructed 1000 curves (or realizations) of unanchored luminosity distance $H_0D^L$ from the two SNe datasets are shown in the Fig.~\ref{fig:sn}, respectively. These curves show the shape of  distance-redshift relation very well.
It should be noted that the redshift of the SNe dataset well covers redshift range of BAO data, so
there is no need for extrapolation of the redshift range. Further, based on the Etherington relation,
$D^A=D^L/(1+z)^2$, we convert the 1000 realizations $H_0D^L$ to $H_0D^A$ served for BAO data.

\section{Results and Discussion}
\begin{figure*}
\begin{center}
{\includegraphics[width=0.48\linewidth]{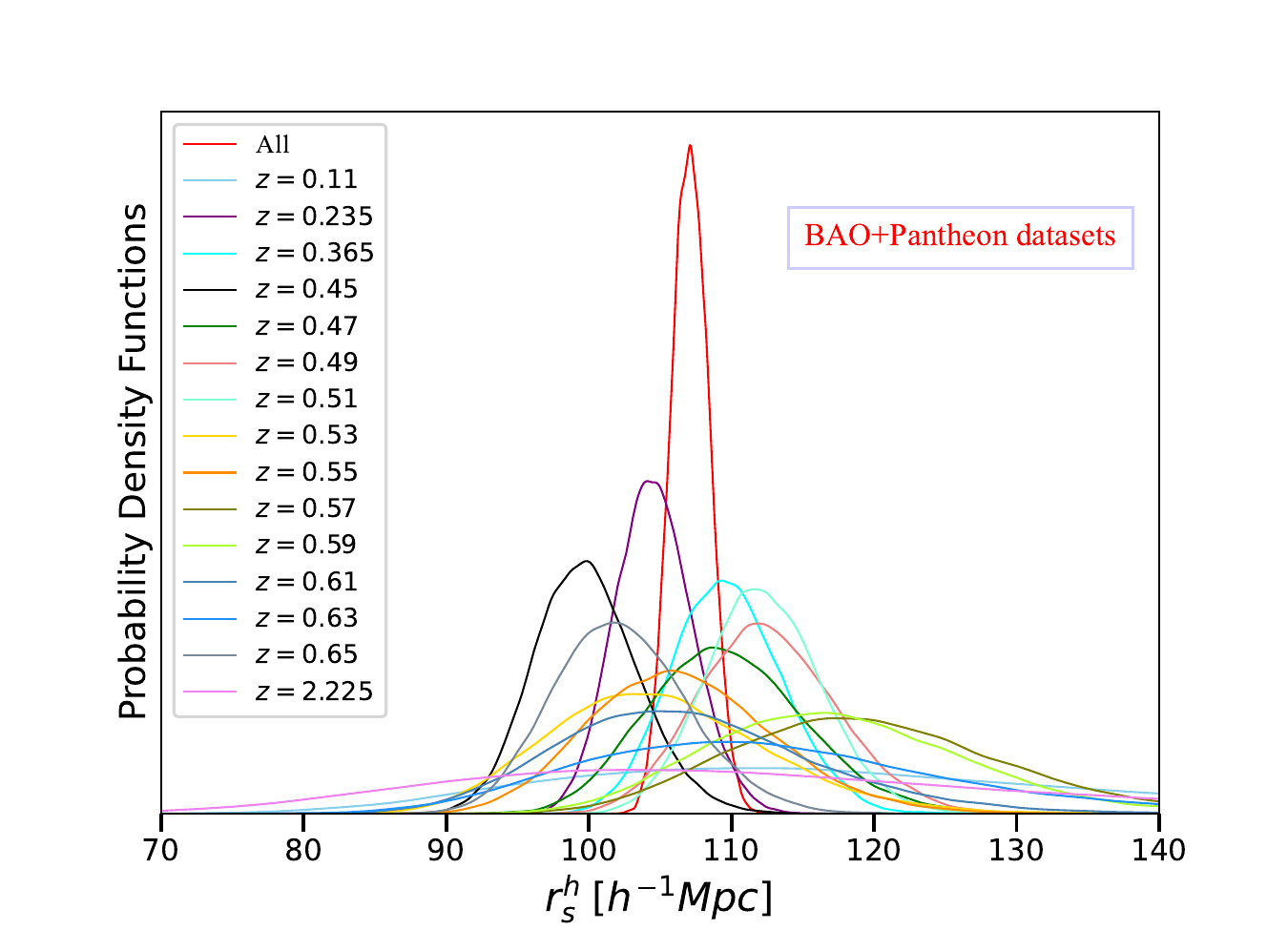}
\includegraphics[width=0.48\linewidth]{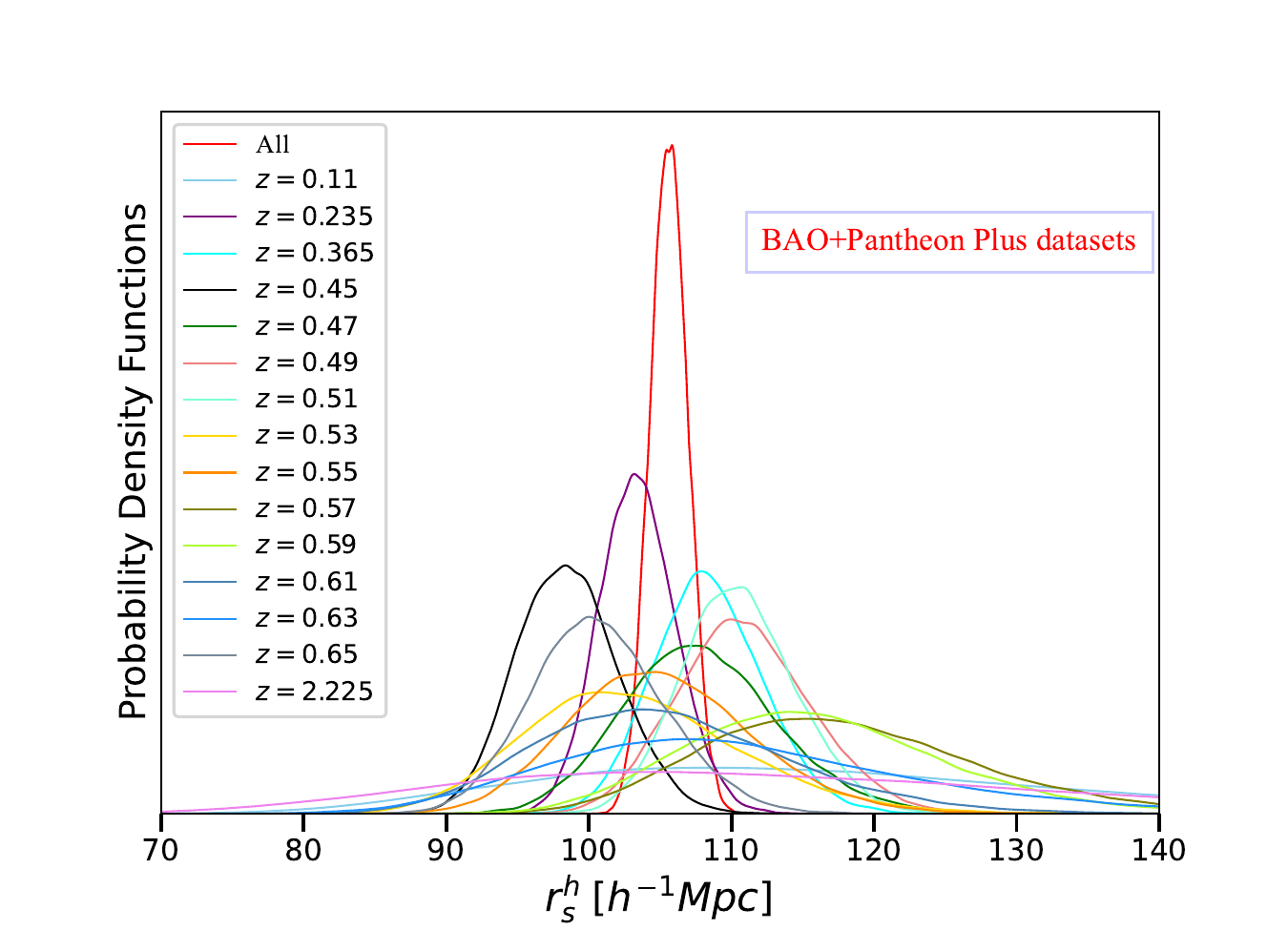}}
\end{center}
\caption{ {\it Left panel:} The probability distribution functions for the relative standard ruler $r_s^{h}$ using transversal BAO plus Pantheon datasets. {\it Right panel:} Similar to left figure, but for Pantheon+ dataset.}\label{fig2}
\end{figure*}

%\subsection{Methodology for calibration of BAO relative standard ruler $r_s^{h}$}
Let's emphasize that calibration of BAO relative standard ruler $r_s^{h}$ using observations of SNe Ia is straightforward, we do not make assumptions about cosmological models and other cosmological parameter.
The BAO relative standard ruler $r_s^{h}$ can be rewritten as
\begin{equation}\label{BAODA}
r_s^h=\frac{(1+z_{BAO})\theta_{BAO}}{100}\cdot [H_0D^A]^{SNe}.
\end{equation}
By combining 15 transverse BAO measurements and reconstructed 1000 realizations $H_0D^A$ from SNe datasets, the individual Probability Distribution Functions (PDFs) for $r_s^{h}$ are shown displayed in Fig. \ref{fig2}. The uncertainty on $[H_0D^A]^{SNe}$ and $\theta_{BAO}$, as well as their mutual correlation are intrinsically included in the respective PDFs. Then, we use these samples to reconstruct the multi-variate distribution of the 15 $r_s^{h}$ measurements.

\begin{figure}
\begin{center}
\includegraphics[width=7.5cm,height=6.3cm]{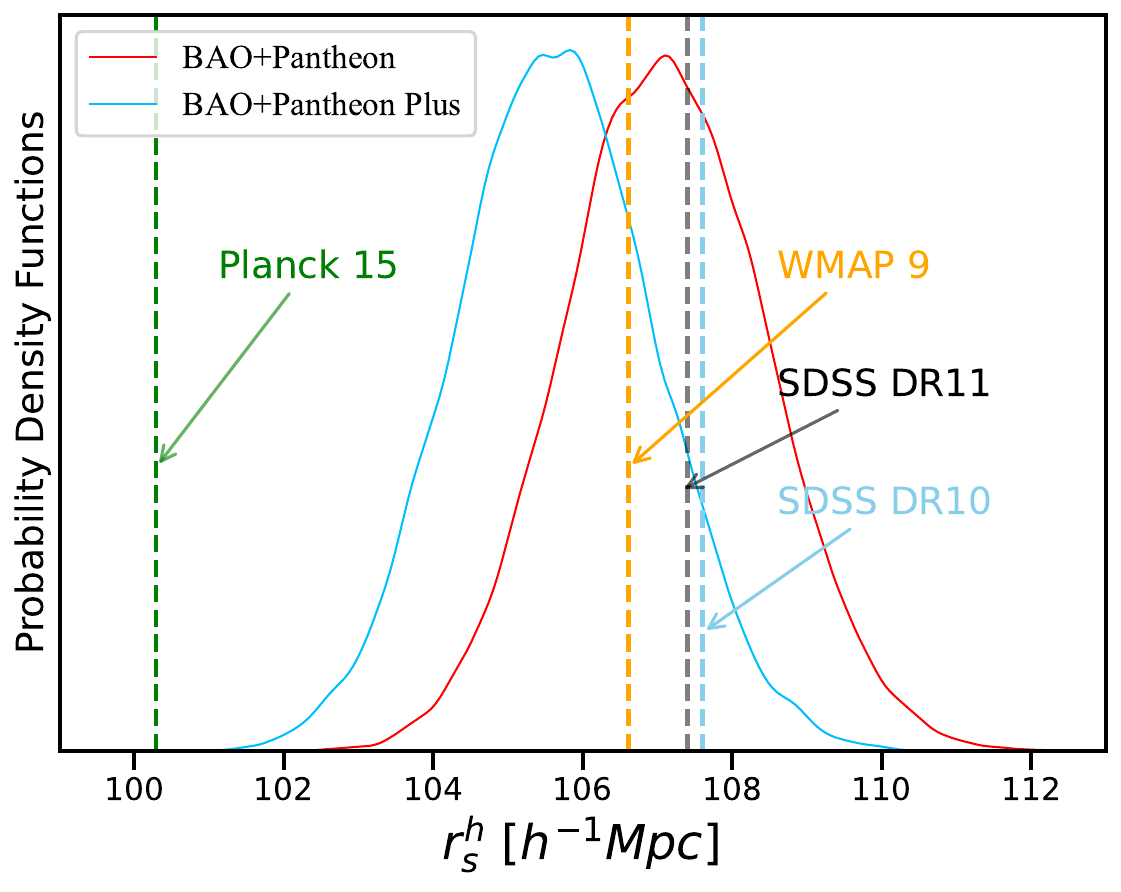}
\end{center}
\caption{The probability density function of $r_s^{h}$ using all transversal BAO measurements and two SNe Ia datasets and comparing our results with previous works.}\label{fig3}
\end{figure}

\begin{figure}
\begin{center}
\includegraphics[width=8.5cm,height=6.3cm]{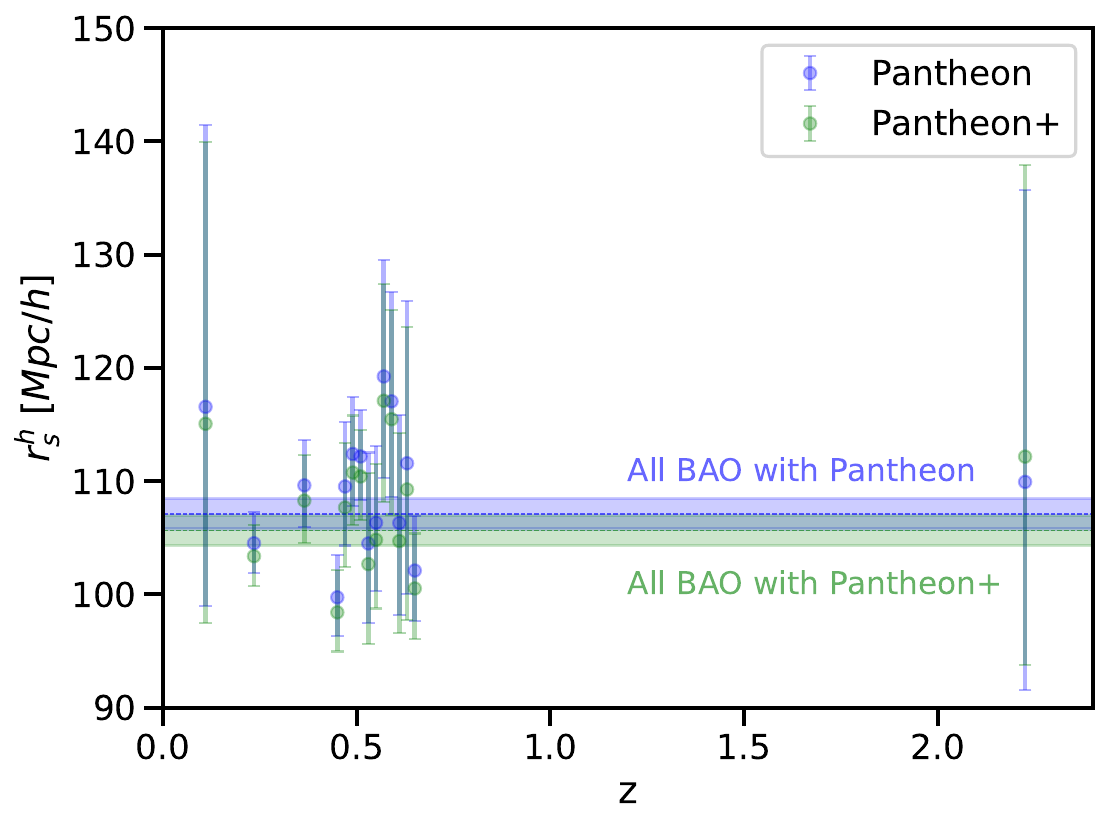}
\end{center}
\caption{The 15 independent $r_s^{h}$ calibration results combined transversal BAO measurements with two SNe Ia datasets.}\label{fig4}
\end{figure}

We note that our results for $r_s^{h}$ are individual results, which means that we achieve calibrations of $r_s^{h}$ at different redshifts. Work on the Pantheon dataset, the final result obtained combining these individual PDFs reports  $r_s^{h}=107.10^{+1.36}_{-1.32}$ $Mpc/h$ (median value plus the $16^{th}$ and $84^{th}$ percentiles around this).  In order to highlight the potential of our method, it is necessary
to compare our results with previous works. Assuming the standard cosmology with matter density parameter $\Omega_m=0.27$, the Nine-Year Wilkinson Microwave Anisotropy Probe (WMAP 9) collaboration reported the value of sound horizon $r_s^{h}=106.61\pm3.47$ $Mpc/h$ \cite{2013ApJS..208...19H}, and Planck Collaboration 2015 (Planck 15) observation derived  $r_s^{h}=100.29\pm2.26$ $Mpc/h$ \cite{2016A&A...594A..13P}. Other work such as Refs \cite{2016PhRvD..93b3530C,2020APh...11902432C} used the SDSS data release 10 (SDSS DR10)  and  11 (SDSS DR11) galaxies with a prior of matter density parameter to obtain $r_s^{h}=107.6\pm2.3$ $Mpc/h$ and  $r_s^{h}=107.4\pm1.7$ $Mpc/h$, respectively. The work \cite{2017MNRAS.467..731V} also obtained constraints on the length of the low-redshift standard ruler $r_s^{h}=107.4\pm3.4$ $Mpc/h$ when using standard clocks+BAO measurements with the local $H_0$ prior.  It is important to note that these work either assumes cosmological models or uses prior values, whereas our calibration of $r_s^{h}$ comes only from observations. Nevertheless, compared with previous work, our calibrated $r_s^{h}$ results are in agreement with the above results, except for Planck 2015 results, as shown in Fig. \ref{fig3}, and the precision of our results on $r_s^{h}$ is particularly high.
In addition, it is interesting to see from the Fig. \ref{fig4} that most of the $r_s^{h}$ obtained at high redshifts have a larger value (9 out of 15 results are larger than the result obtained by combining all BAOs). Assuming absolute standard ruler $r_s$ is a fixed value, the $H_0$ will decrease when $r_s^{h}$ increases (or redshift increases) based on relation $r_s=100r_s^{h}/H_0$.  This finding may give us a better understanding of the discordance between the data sets or Hubble tension or reveal new physics beyond the standard model.
%An almost model-independent measurement of the BAO ruler length was recently reported in Ref. \cite{Phys. Rev. Lett. 113, 241302}. Using type Ia supernova and galaxy clustering data, the authors obtained $r_s^{h}=101.9\pm1.9$ $Mpc/h$.

Work on the Pantheon+ dataset, our model independent calibrated result is $r_s^{h}=105.63^{+1.33}_{-1.31}$ $Mpc/h$ with median values plus 16$^{th}$ and 84$^{th}$ percentiles and shown in the right panel of Fig. \ref{fig2}.  Comparing the results obtained from the two SNe datasets, we find that the Pantheon+ results are only slightly different from the result using  Pantheon dataset.
This is because the lowest BAO redshift is 0.11. Compared with Pantheon sample, the main change is the sample size of Pantheon+ (especially in $z <0.01$) and a larger redshift span. The Pantheon+ sample comprises 77 data points from Cepheid host galaxies at very low redshifts \cite{2022ApJ...934L...7R,2022ApJ...938..110B}, but remaining or not remaining these data points has little impact on our work. Since GPR is a completely data-driven method,  the results of the reconstruction depend on the input data, and the low redshift data is not used, the reconstruction results based on the two datasets are almost identical, we get almost the same results with the two SNe datasets.
In addition, the dataset using Pantheon+ does not improve the precision of $r_s$ calibration (at the second significant digit). Note that even without an estimate of dimensionless Hubble constant $h$, the combination of BAO and SNe Ia datasets already constrain the low-redshift standard ruler scale $r_s^{h}$ at the $\sim1.26\%$ level.

\section{Conclusion}

In this work, we have proposed a new model-independent method to calibrate sound horizon $r^h_s$ (relative standard ruler) by using the latest observations of SNe Ia acting as standard
candles and transversal BAO measurements.  We adopt the non-parameterized method GPR to reconstruct two SNe Ia datasets, which provide history of the expansion of the universe. Our work has some significant advantages: it has nothing to do with cosmological models, and all the quantities used come from observations; no prior values are taken for any and cosmological parameters, including $H_0$; only two data samples are involved, highlighting the contribution of each dataset to the overall work.

The final result obtained combining all transversal BAO measurements reports  $r_s^{h}=107.10^{+1.36}_{-1.32}$ $Mpc/h$ (median value plus the $16^{th}$ and $84^{th}$ percentiles around this) in the framework of the Pantheon dataset. This result changes to  $r_s^{h}=105.63^{+1.33}_{-1.31}$ $Mpc/h$ when uses Pantheon+ dataset.  Note that even without an estimate of dimensionless Hubble constant $h$, the combination of BAO and SNe Ia datasets already constrain the low-redshift standard ruler scale $r_s^{h}$ at the $\sim1.26\%$ level. More importantly, it is interesting to find that most of the $r_s^{h}$ obtained at high redshifts have a larger value (9 out of 15 results are larger than the result obtained by combining all BAOs).  This finding may give us a better understanding of the discordance between the data sets or Hubble tension or reveal new physics beyond the standard model.

The GPR method we used here is not unique. It is still worth exploring whether these conclusions will change with different reconstruction methods such as machine learning method.
However, these methods have their own advantages and disadvantages, and show great potential in the studies of precision cosmology. For instance, the GPR greatly reduces the uncertainty of data.  Some work has generally used non-parametric GP method to reconstruct the expansion history of the universe. However, as mentioned in many works \cite{2018MNRAS.478.3640M,2022JCAP...12..029M}, one of the most recently debated topics for non-parametric reconstruction in cosmology with GP is that this technique is exposed to several foundational issues such as overfitting and kernel consistency problems \cite{2021arXiv210108565C}.  GPR is different from GP, since the GPR occurs in an infinite-dimensional function space without overfitting problem. Meanwhile, a recent work \cite{2023ApJS..266...27Z} indicated that using different kernels would not affect the GP reconstruction significantly. Before the era of abundant available data, it is also very necessary to carefully choose the method of data reconstruction.

As a final remark, the precision of calibration $r_s$ is currently constrained mainly by samples of BAO data. The first is the BAO sample size, followed by the observed error of the angle scale. The arrival of the next generation of BAO surveys has great potential to improve low redshift standard scale measurements. For example, the Dark Energy Spectroscopic Instrument (DESI) collaboration recently release the first year data
(DR1) of baryon acoustic oscillations (BAO) to substitute the current BAO measurements \cite{2024arXiv240403000D,2024arXiv240403001D}, thereby may directly or indirectly affecting $H_0$ measurements. This shows that our method has great potential to provide more precise and accurate measurements of $r_s$ in the future, further precision cosmology research.

\section*{Acknowledgments}
Liu. T.-H was supported by National Natural Science Foundation of China under Grant No. 12203009;  Chutian Scholars Program in Hubei Province (X2023007); Marek. B. was supported by the  Hubei Province Foreign Expert Project (2023DJC040).
Wang. J. C  was supported by the National Natural Science Foundation of China under Grants No. 12122504 and No. 12035005; the innovative research group of Hunan Province under Grant No. 2024JJ1006; the Natural Science Foundation of Hunan Province under grant No. 2023JJ30384; and the Hunan Provincial Major Sci-Tech Program under grant No.2023ZJ1010.

\end{document}